\documentclass[%
 reprint,
 amsmath,amssymb,
 aps,
]{revtex4-2}
\usepackage{mathtools}
\usepackage{braket}
\usepackage{graphicx}% Include figure files
\usepackage{dcolumn}% Align table columns on decimal point
\usepackage{bm}% bold math
\begin{document}

\preprint{APS/123-QED}

\title{Splitting indistinguishable photons: Using linear optics to exceed the limit of photon blockade }% Force line breaks with \\

\author{Harjot Singh}
 \email{hsingh16@umd.edu}
\author{Edo Waks}%
 \email{edowaks@umd.edu}
\affiliation{%
 Department of Electrical and Computer Engineering,
 University of Maryland, College Park, Maryland, USA %\textbackslash\textbackslash
}%


%\affiliation{
%\affiliation{
%\affiliation{%


\date{\today}% It is always \today, today,

\begin{abstract}
Photon-photon interactions are an essential requirement of quantum photonic information processing. One way to generate these interactions is to utilize an atom strongly coupled to an optical cavity. This system exhibits the photon blockade effect which enables single photon switching and creation of non-classical light.  But the nonlinear effects enabled by this system suffer from a fundamental time-bandwidth constraint. For the the simple case of splitting an input pulse of two indistinguishable photons, this constraint imposes a limit on the efficiency of routing photons to different output ports.  We show that this limit can be exceeded by combining the strongly-coupled atom with linear optics. By optimizing the unitary of the linear optical transformation, we achieve improved splitting efficiency for both un-entangled and entangled photons. Our results suggest that it may be possible to improve the efficiency of nonlinear optical processes at the single photon level by making suitable use of linear optics. These results could have implications for quantum information processing with photons. 
%\begin{description}
%\end{description}
\end{abstract}

\maketitle

%\tableofcontents

\section{\label{sec:level1} Introduction}

Photon-photon interactions are essential for photonic quantum information processing ~\cite{cirac1997quantum, kok2007linear}. They enable two qubit gates between photons, which are a key requirement for quantum computation ~\cite{duan2004scalable}. They also play an important role in photonic quantum simulation \cite{aspuru2012photonic}. But generating these interactions is challenging because they require optical nonlinearities at the single photon level.      

One way to generate optical nonlinearities at the single photon level is to      exploit strong light-matter interactions between a two-level atom and an optical cavity \cite{kimble1998strong}. These systems can exhibit photon blockade \cite{birnbaum2005photon}, which allows a single photon to strongly modify the cavity reflectivity and thus      control the transmission and reflection of subsequent photons~\cite{haroche1989cavity,birnbaum2005photon, faraon2008coherent}. If photon blockade shown by an atom-cavity system were ideal,  two indistinguishable photons incident on the system would be split to different output ports, something which cannot be done with linear optics with probability greater than $50\%$\cite{rosenblum2011photon}.  But all previous realizations of photon blockade failed to achieve a complete splitting of two indistinguishable photons due to a fundamental time-bandwidth limit, originally      analyzed by Rosemblum et. al. \cite{rosenblum2011photon}.  This limit imposes a maximum splitting efficiency of      66\% for an input pulse containing unentangled photons, and 77\% when time-energy entanglement is introduced between input photons. This limit constrains the achievable single photon nonlinearity using a two-level atom in a cavity\cite{nysteen2017limitations}.

In this work, we show that this limit in splitting efficiency can be exceeded by adding linear optics after the atom-cavity system. These linear optics create a unitary transformation between the output ports of the atom-cavity system.  By      optimizing      the unitary transformation      we achieve      a splitting efficiency of greater than 75\% for unentangled input photons, and 90\% for entangled input photons. The improvement in splitting efficiency is enabled by two-photon interference between the two output ports of the atom-cavity system, which we exploit through the optimized unitary.  Our results suggest that it may be possible to extract stronger nonlinear effects from a two-level atom than previously believed.

\section{Splitting efficiency for unentangled photons}
Fig. 1 shows the experimental setup we analyze. The photon enters the input port of a microtoroidal cavity, defined in the diagram as $\hat{a}_{in}$.     The microtoroidal cavity has both a clockwise and counter-clockwise propagating mode, each of which couple to two-level atom with equal cavity-atom coupling constants $g$.  The counterclockwise mode decay to output mode $\hat{a}_{out}$ with a decay rate of $2\kappa_{ex}$, while the clockwise mode decays to mode bout with the same rate.   This part of the system is identical to the one originally studied by Rosenblum et. al., and is known to have a limited routing efficiency.
\begin{figure}
\includegraphics[scale=0.75]{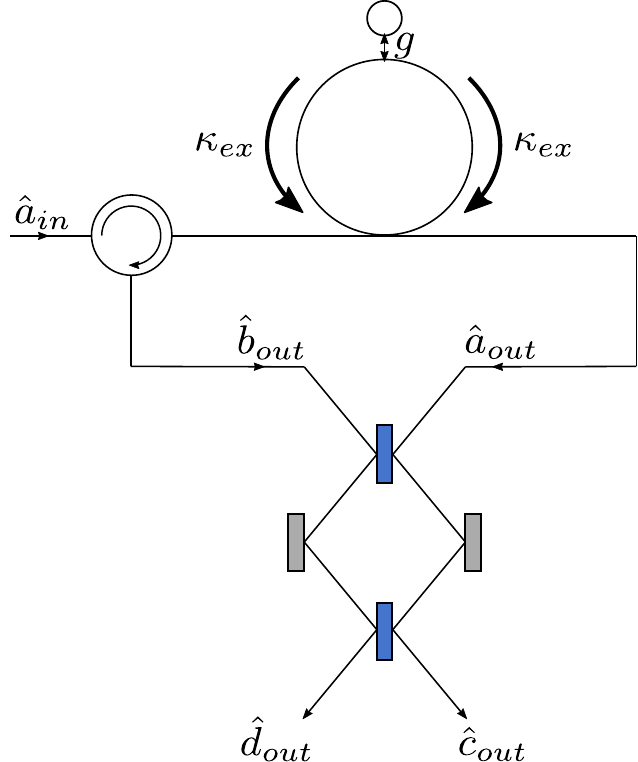}% Here is how to import EPS art
\caption{\label{fig:wide}Our setup consists of an atom coupled to a microtoroidal cavity, a Mach Zehnder Interferometer and a fiber port circulator. A 2 photon input is sent from $\hat{a}_{in}$. The fiber port circulator sends the light reflected from the cavity to $\hat{b}_{out}$. The interferometer implements the most general unitary operation on $\hat{a}_{out}$ and $\hat{b}_{out}$}
\end{figure}

In order to improve the splitting efficiency we inject the output ports of the cavity into an arbitrary linear optical unitary   operation, as shown in Fig. 1.  We restrict our attention specifically to unitaries with two-input and two-output modes to preserve the same photonic Hilbert space. In this case, the Mach-Zehnder interferometer shown in Fig. 1 can apply the most general linear optical unitary on the photonic states. The Mach-Zehnder interferometer transforms the modes $\hat{a}_{out}$ and $\hat{b}_{out}$ into modes $\hat{c}_{out}$ and $\hat{d}_{out}$  according to the transformation:

\begin{equation}
\left(
\begin{array}{c}
\hat{c}_{out} \\
\hat{d}_{out}
\end{array}\right) = 
\begin{pmatrix}
e^{i\phi}sin(\omega) & cos(\omega) \\
e^{i\phi}cos(\omega) & -sin(\omega)
\end{pmatrix}
\begin{pmatrix}
\hat{a}_{out} \\
\hat{b}_{out}
\end{pmatrix}
\label{eq:one}
\end{equation}

In the above equation the unitary is controlled by two phase parameters, $\omega$ and $\phi$. The phase $\omega$  controls the amplitude splitting ratio of the Mach Zehnder interferometer, and physically corresponds to the path length difference between the two interferometer arms.  The other phase parameter, $\phi$, applies an additional phase shift between the two input modes of the interferometer. Our goal is to optimize the two parameters of the unitary to improve photon-photon splitting.

The figure of merit that quantifies how close the system behaves to an ideal single photon splitter is the splitting efficiency. We define this figure as the probability of detecting one photon each in the output modes $\hat{c}_{out}$ and $\hat{d}_{out}$, given that the input mode $\hat{a}_{in}$ contains two photons initially. We calculate the splitting efficiency from the two-time correlation functions.   

\begin{eqnarray}
\Gamma^{cc} = \left<\hat{c}^\dagger_{out}(t)\hat{c}^\dagger_{out}(t+\tau)\hat{c}_{out}(t+\tau)\hat{c}_{out}(t)\right> \label{eq:two}
\\
\Gamma^{dd} = \left<\hat{d}^\dagger_{out}(t)\hat{d}^\dagger_{out}(t+\tau)\hat{d}_{out}(t+\tau)\hat{d}_{out}(t)\right> \label{eq:three}
\\
\Gamma^{cd} = \left<\hat{c}^\dagger_{out}(t)\hat{d}^\dagger_{out}(t+\tau)\hat{d}_{out}(t+\tau)\hat{c}_{out}(t)\right> \label{eq:four}
\\
\Gamma^{dc} = \left<\hat{d}^\dagger_{out}(t)\hat{c}^\dagger_{out}(t+\tau)\hat{c}_{out}(t+\tau)\hat{d}_{out}(t)\right> \label{eq:five}
\end{eqnarray}

The correlation function $\Gamma^{cd}$ is the probability density of detecting a photon in mode cout at time t, and then a second photon in mode dout  at time $t+\tau$, while $\Gamma^{cd}$ is similarly defined with the two output modes reversed.  The probability densities of detecting the two photons at the same port, $\Gamma^{cc}$ and $\Gamma^{dd}$, can be defined in the same way. We integrate the correlations in Eqs. (4) and (5) over both t and $\tau$ to obtain the total probabilities:

\begin{eqnarray}
P_{cd}=\int_0^\infty\int_0^\infty \!\Gamma^{cd}(t,\tau)  \, \mathrm{d}t\mathrm{d}\tau \\
P_{dc}=\int_0^\infty\int_0^\infty \!\Gamma^{dc}(t,\tau)  \, \mathrm{d}t\mathrm{d}\tau
\end{eqnarray}
The splitting efficiency, denoted $S$, is the sum of these two probabilities:  $S=P_{cd}+P_{dc}$. 

Fig. 2 illustrates the numerical model we use to analyze the system in Fig. 1.  We perform our analysis in the fast cavity regime which allows us to adiabatically eliminate the cavity modes and approximate the atom-cavity system as a 1D atom that couples to the modes of the fiber with an enhanced decay rate $2\gamma=2g^2/\kappa$ \cite{walls2007quantum}.  To model a quantum light source, we employ a feeder cavity      at the input to the atom-cavity system. We define $\hat{a}$ as the bosonic operator for the feeder cavity, and $\hat{\sigma}$ the atomic lowering operator for the effective 1D two-level atom. The output of the feeder cavity connects to the input of the 1D atom such that $\hat{a}_{in} = \sqrt{2\kappa}\hat{a}$, where $2\kappa$ is the decay rate of the feeder cavity. This decay rate controls the resulting photon pulse width, such that a decay rate of zero corresponds to a monochromatic input, and the pulse bandwidth increases monotonically with the decay rate.  We can calculate the dynamics of the system in Fig. 2 using the non-Hermitian Hamiltonian ~\cite{rosenblum2011photon, carmichael1993quantum}. 

\begin{equation}
H_0 = -i\left(\kappa\hat{a}^\dagger\hat{a} + 2\gamma\hat{\sigma}^\dagger\hat{\sigma} + 2\sqrt{\kappa\gamma}\hat{\sigma}^\dagger\hat{a}\right)
\end{equation}

This non-Hermitian Hamiltonian, combined with quantum jumps, fully describes the evolution of the quantum system ~\cite{haroche2006exploring, carmichael2009statistical}.  Here, the quantum jumps are generated by the collapse operators $\hat{a}_{out} = \sqrt{2\kappa}\hat{\sigma} + \sqrt{2\kappa}\hat{a}$ and
$\hat{b}_{out} = \sqrt{2\kappa}\hat{\sigma}$ ~\cite{carmichael1993quantum, gardiner1985input, gardiner1993driving, gardiner1994driving}.

Each of these collapse operators corresponds to a photon detection event at the respective output port, which removes a quantum excitation from the system. We don't need to consider contributions from vacuum noise operators $\hat{a}_{s,in}$ and $\hat{b}_{in}$ to these collapse operators because the correlations we calculate in Eq. 2-5 have the field operators in normal order \cite{rosenblum2011photon}.

We first analyze the splitting efficiency for the case where the feeder cavity initially contains a two-photon fock state. This initial condition generates an output field of two indistinguishable photons that are unentangled in time and frequency.  Appendix A provides the full derivation of the splitting efficiency after the unitary transformation. The splitting efficiency is given by: 
\begin{widetext}
\begin{equation}
S = \frac{8 \gamma ^3+16 \gamma ^2 \left(\sin ^2(2 \omega ) \cos (2 \phi )+2 \sin (4 \omega ) \cos (\phi )\right)+44 \gamma ^2+(2 \gamma  (2 \gamma  (5-2 \gamma )+5)-3) \cos (4 \omega )+38 \gamma +3}{4 (2 \gamma +1)^2 (2 \gamma +3)}
\end{equation}
\end{widetext}

Fig. 3 shows the resulting calculated splitting efficiency      as a function of   and $\frac{\gamma}{\kappa}$, with $\phi$ set to 0 because, as shown in the appendix A1, we end up not requiring a phase difference between the input ports of the Mach-Zehnder interferometer to optimize the splitting efficiency. Since the splitting efficiency is periodic with $\omega$, we plot only one period.       The splitting efficiency      achieves a maximum of 75\% at $\frac{\gamma}{\kappa}=0.92$, and $\omega=0.303$ . Fig. 3b plots a cross-section of Fig. 3a for two different unitary transformations.  The first corresponds to the optimal transformation given by $\omega=0.303$. The second is $\omega=0$ which is the identity transformation, corresponding to the case where we remove the Mach-Zehnder interferometer and look directly at the output modes of the atom.  In this case           we achieve the same maximum splitting efficiency of 64\% previously derived in Ref. \cite{rosenblum2011photon}.  The optimized unitary therefore improves the splitting efficiency over the limit implied by the time-bandwidth constraint, which was derived in Ref. \cite{rosenblum2011photon}.  
\begin{figure}
\includegraphics[scale=0.75]{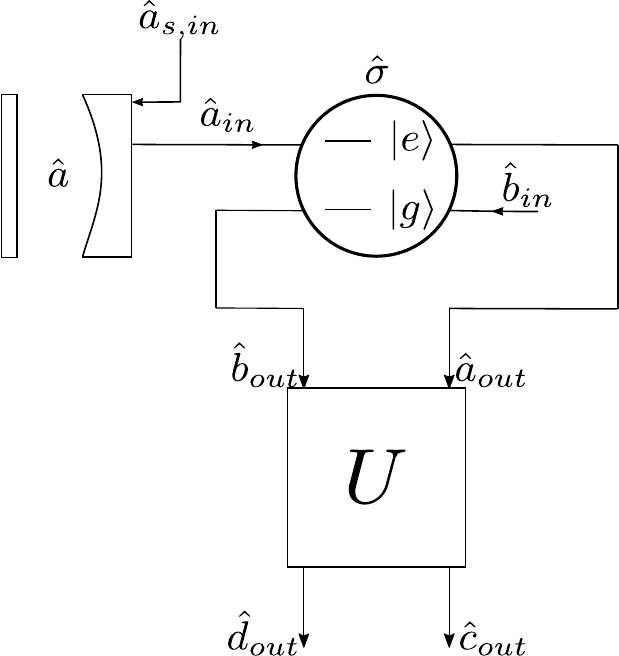}% Here is how to import EPS art
\caption{\label{fig:wide}Schematic of the model we analyse. A feeder cavity is the source of a 2 photon pulse. Atom-cavity system is replaced with an enhanced atom after adiabatic elimination of cavity modes. The box with the letter U denotes a unitary transformation. }
\end{figure}
\begin{figure*}
\includegraphics[scale=0.6]{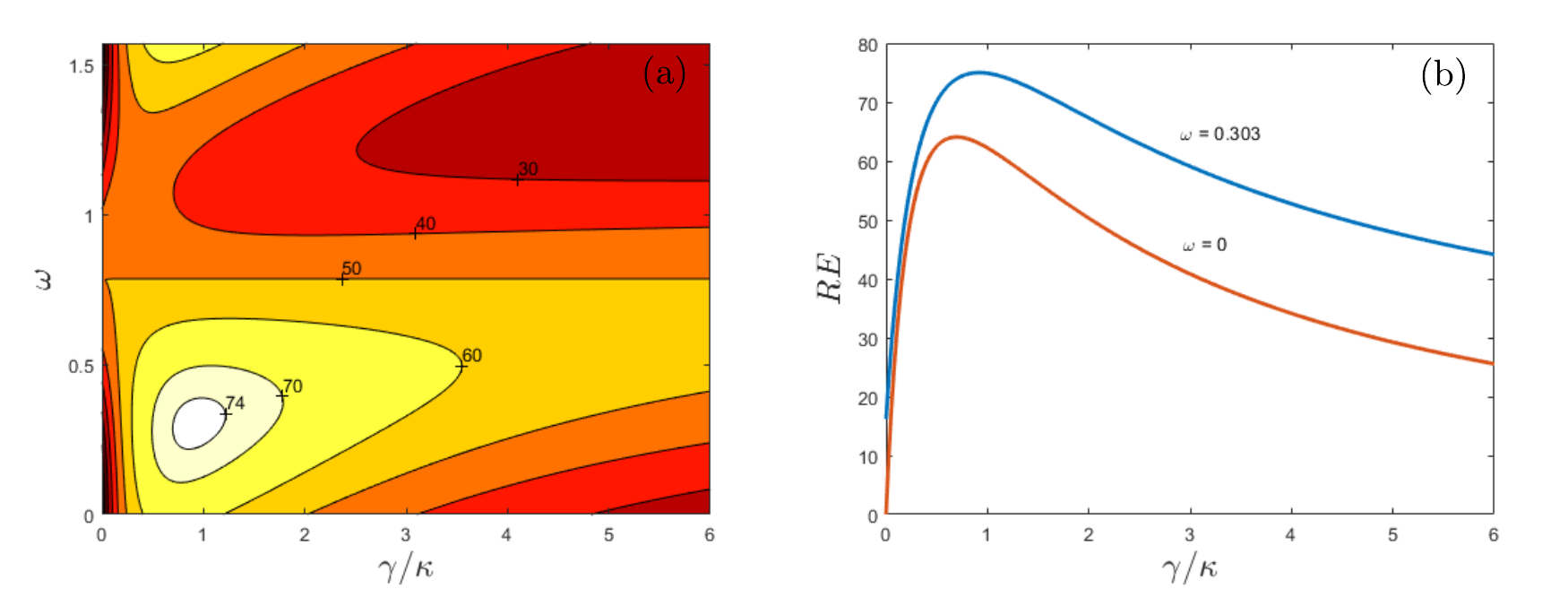}% Here is how to import EPS art
\caption{\label{fig:wide}(a) Contour plot of routing efficiency for an input of 2 unentangled photons. (b) Slices of the contour plot corresponding to the splitting efficiencies for the optimal unitary transformation at $\omega = 0.303$ and the identity matrix.}
\end{figure*}

To better understand why a      linear optical transformation can improve the splitting efficiency, we apply a simplified single-mode model for the output of the 1D atom.  We assume the output state of the two-level atom, which is input to a Mach Zehnder Interferometer (Fig. 1), is given by $\ket{\Psi_{ab}} = d\ket{11} + e\ket{20} + f\ket{02}$, where the state $\ket{nm}$ corresponds to the case where there are $n$ photons in mode $\hat{a}_{out}$ and $m$ photons in $\hat{b}_{out}$. In this simple case, the routing efficiency is the probability of finding $\ket{\Psi_{cd}} = \ket{11}$ state at the output of the Mach Zehnder interferometer.  We can rewrite the input to the interferometer as     $ \ket{\psi_{ab}} = d  \ket{11}+\left(e+f\right)/\sqrt{2}\ket{+}+(e-f)/\sqrt{2}\ket{-}$, where $\ket{+} = \ket{20} + \ket{02}$ and $\ket{+} = \ket{20} - \ket{02}$. The unitary operation realized by the interferometer transforms the input states $\ket{-}$  and $\ket{11}$ into the states $cos(2\omega)\ket{11} + sin(2\omega)\ket{-}$ and $sin(2\omega)\ket{11} + cos(2\omega)\ket{-}$  at the output of the interferometer respectively,
while leaving state $\ket{+}$ unchanged.  These transformations represent generalized two-photon interference for a beam splitter with any splitting ratio      (see Appendix B) .  The Hong-Ou-Mandel interference is a special case  where the input state is $\ket{11}$ and the unitary creates 50/50 beamsplitting.       If the complex amplitudes $d$, $e$ and $f$ have the same phase and $e = -f = g/\sqrt{2}$,such that the input to the interferometer is of the form      $ \ket{\psi_{ab}} = d  \ket{11}+ g \ket{-}$, a      linear optical transformation can always generate the required two-photon interference effect that transforms the state      to $\ket{11}$ at the output, giving 100\% routing efficiency.       However, we can’t generate this ideal input state with our 1D atom (figure 2),  due to photon antibunching in the $\hat{b}_{out}$  mode, which means $f=0$ (See Appendix B) . This prevents the achievement of ideal routing efficiency.

\section{Splitting efficiency for Time-energy entangled photons}
The routing efficiency calculated in the previous analysis could potentially be improved by introducing time-energy entanglement between the two photons.  Indeed,      Rosenblum et. al. \cite{rosenblum2011photon} showed that a time-energy entangled input state      could improve the optical routing efficiency from 64\% to      77\%. By introducing an optimized linear-optical transformation we could potentially improve this routing efficiency even further.

To incorporate time-energy entanglement into our analysis, we utilize the model introduced by Rosenblum et. al. which is illustrated in      Figure 4     .  Instead of initializing the feeder cavity to a two-photon fock state, we instead excite the cavity with a three-level atom with equally spaced energy levels. The intermediate level has a short lifetime, resulting in a pair of photons with time-energy entanglement. The Hamiltonian for this cascaded system is given by:

\begin{eqnarray}
  H_1=-i(\hat{\kappa}{\hat{a}}^\dag\hat{a}+2\gamma {\hat{\sigma}}^\dag\hat{\sigma} 
  +2\sqrt{\kappa\gamma }{\hat{\sigma}}^\dag\hat{a}+\nonumber\\2\delta {\hat{\sigma}}_s^\dag{\hat{\sigma}}_s + 2\sqrt{2\chi\delta }\left({\hat{a}}^\dag\right)^2{\hat{\sigma}}_s)  
\end{eqnarray}

In the above equation,  $\hat{\sigma}_s$ is the lowering operator  of the three-level atom from the upper state $\ket{e}$ to the ground state $\ket{g}$ and it decays with a rate of $4\delta$. $\chi$ is the bandwidth of the left mirror of the feeder cavity and is assumed to be very small \cite{rosenblum2011photon}. 

 \begin{figure}
\includegraphics[scale=0.65]{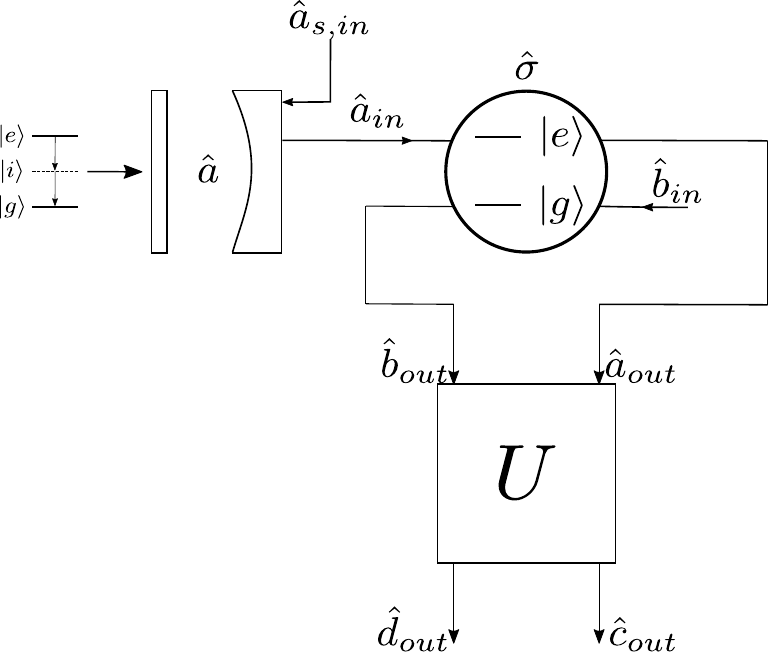}
\caption{\label{fig:wide}Model for calculating the splitting efficiency for a pair of time-energy entangled photons. A three level atom with two cascaded transformations with identical frequencies generated the entangled photon pair which is input to the feeder cavity. The rest of the model is the same as in fig. 2.}
\end{figure}

We employ the same approach as for the unentangled photons to analyze the routing efficiency. We initialize the three-level atom in state $\ket{e0g}$, which contains two excitations, while the cavity is empty, and the two-level atom starts in the ground state. We follow the same steps as in Appendix A to obtain an analytical expression for splitting efficiency and equation C1 gives the expression for splitting efficiency in this case.

 \begin{figure}
\includegraphics[scale=0.55]{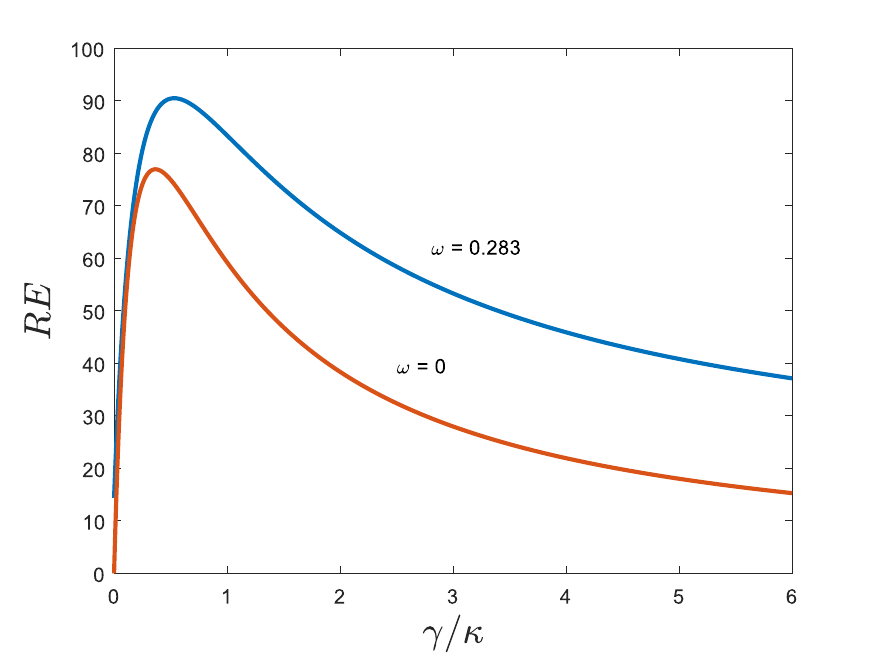}% Here is how to import EPS art
\caption{\label{fig:wide}Splitting efficiency for an input of time-energy entangled photons. $\omega=0.283$ gives the optimal unitary transformation and $omega=0$ corresponds to the splitting efficiency for the identity transformation. Here, $\kappa/\delta$ approaches infinity, which corresponds to maximum time-energy entangled between the input photons.}
\end{figure}
Figure 5 plots the resulting splitting efficiencies for a      time-energy entangled two-photon state. We plot the splitting efficiencies for two unitary transformations corresponding to $\omega=0.283$, which results in the maximum routing efficiency, and $\omega = 0$, which gives the splitting efficiency for when the interferometer is absent from the setup in Fig. 1. The two results are plotted in the limit $\kappa/\delta \xrightarrow[]{}$ $\infty$, which corresponds to maximum time-energy entanglement between input photons and gives the highest splitting efficiency in both cases. When the unitary transformation corresponds to the optimal case, the maximum splitting efficiency is 90\%.   If we don't perform a unitary transformation of the modes $\hat{a}_{out}$ and $\hat{b}_{out}$,  the maximum achievable splitting efficiency is 77\%. Thus, by tuning the unitary transformation, we improve the splitting efficiency for a pair of time-energy entangled photons. 

In      summary, we      showed that  linear optics      can improve on the splitting efficiency possible with photon blockade using an atom-cavity system.      We improved the splitting efficiency from 64\% to 75\% for unentangled photons, and      from 77\% to 90\% for      photons that      are time-energy entangled. We showed how 2 photon interferences created by the unitary transformation realized by a Mach Zehnder Interferometer enables this increase in the splitting efficiency. By changing the temporal nature of the interference, i.e. by varying the unitary transformation in time could lead to a further improvement. 2 photon controlled phase gate implemented with the nonlinearity of a two-level atom also suffers from the time-bandwidth limit in its fidelity \cite{nysteen2017limitations}. Incorporating interference of the output modes of the two-level atom with a general linear optical transformation may lead to an increase in the gate fidelity.  Our results highlight the potential to generate improved quantum optical interaction by combining      single-photon nonlinearities with linear optical systems, with potential applications in photon quantum information processing and quantum simulation.

\begin{acknowledgments}
The authors would like to acknowledge financial support from the National Science Foundation (grants \#OMA1936314 and \#ECCS1933546), and AFOSR grant \#FA23862014072, and the Maryland-ARL Quantum Partnership (W911NF1920181).

\end{acknowledgments}

\appendix

\section{Derivation of splitting efficiency for unentangled photons}
The system in figure 2 is initialized to the state $\ket{2g}$, with two photons in the feeder cavity and the atom in the ground state. Before the detection of the first photon, this state evolves under the Hamiltonian $H_0$ \cite{rosenblum2011photon} to
\begin{equation}
    \ket{\phi(t)} = \alpha(t)\ket{2g} + \beta(t)\ket{1e}
\end{equation}
with
\begin{equation}
\begin{aligned}
\alpha(t) &= e^{-2\kappa t} \\
\beta(t) &= -\frac{2\sqrt{2\gamma\kappa}}{2\gamma-\kappa} (e^{-\kappa t} - e^{-2\gamma t}  ) e^{-\kappa t}
\end{aligned}
\end{equation}

After the first photon is detected, the state of the system collapses to $\ket{1g}$ or $\ket{0e}$. In the first case, the state of the system evolves to
\begin{equation}
    \ket{\phi(t)} = a(t)\ket{1g} + b(t)\ket{0e}
\end{equation}
with
\begin{equation}
\begin{aligned}
a(t) &= e^{-\kappa t} \\
b(t) &= -\frac{2\sqrt{\gamma\kappa}}{2\gamma-\kappa} (e^{-\kappa t} - e^{-2\gamma t}  )
\end{aligned}
\end{equation}

Substituting these expressions into Eqs. 4 and 5 gives the following second order correlations.
\begin{widetext}
\begin{eqnarray}
\Gamma^{cd}(t,\tau) = 4 | c(\tau ) \exp (-i \phi ) (\exp (-i \phi ) \cos (\omega )-\sin (\omega )) \sin (\omega ) \sqrt{\kappa  \gamma } \beta (t)+  \nonumber \\ \left(b(\tau ) (\exp (-i \phi ) \cos (\omega )-\sin (\omega )) \sqrt{\gamma }+  a(\tau) \cos (\omega ) \exp (-i \phi ) \sqrt{\kappa }\right) \nonumber \\ \left(\exp (-i \phi ) \sin (\omega ) \sqrt{2 \kappa }  \alpha (t)+(\cos (\omega )+\exp (-i \phi ) \sin (\omega )) \sqrt{\gamma } \beta (t)\right)| ^2   
\end{eqnarray}
\begin{eqnarray}
\Gamma^{cd}(t,\tau) = 4 | c(\tau ) \cos (\omega ) \exp (-i \phi ) (\cos (\omega )+\exp (-i \phi ) \sin (\omega )) \sqrt{\kappa  \gamma } \beta (t)+\nonumber \\\left(b(\tau ) (\cos (\omega )+\exp (-i \phi ) \sin (\omega )) \sqrt{\gamma }+a(\tau) \exp (-i \phi ) \sin (\omega ) \sqrt{\kappa }\right) \nonumber \\ \left(\cos (\omega ) \exp (-i \phi ) \sqrt{2 \kappa } \alpha (t)+(\exp (-i \phi ) \cos (\omega )-\sin (\omega )) \sqrt{\gamma } \beta (t)\right)| ^2   
\end{eqnarray}
\end{widetext}

Integrating these second order correlations over $t$ and $\tau$ and summing the two results gives the splitting efficiency.
\begin{widetext}
\begin{equation}
S = \frac{8 (\frac{\gamma}{\kappa})^3+16 (\frac{\gamma}{\kappa})^2 \left(\sin ^2(2 \omega ) \cos (2 \phi )+2 \sin (4 \omega ) \cos (\phi )\right)+44 (\frac{\gamma}{\kappa})^2+(2 \frac{\gamma}{\kappa}  (2 \frac{\gamma}{\kappa}  (5-2\frac{\gamma}{\kappa} )+5)-3) \cos (4 \omega )+38\frac{\gamma}{\kappa} +3}{4 (2\frac{\gamma}{\kappa} +1)^2 (2\frac{\gamma}{\kappa}+3)}
\end{equation}
\end{widetext}

$S$ is periodic in $\omega$ and $\phi$ and is a convex function within one period. Using standard calculus, we can obtain a global maximum of $S$ and it occurs at $\phi=0$, $\omega=0.303$ and $\frac{\gamma}{\kappa} = 0.92$.

\section{Two photon interference}

A general single mode input to the Mach Zehnder Interferometer (Figure 2) is given by:

\begin{equation}
\ket{\phi}_{ab} = d\ket{11}_{ab} + e\ket{20}_{ab} + f\ket{02}_{ab}
\end{equation}

To obtain the state at the output of the interferometer for this input, we need to use the inverse of the unitary transformation given by eq. 2., which is
\begin{equation}
\left(
\begin{array}{c}
\hat{a}^\dagger_{out} \\
\hat{b}^\dagger_{out}
\end{array}\right) = 
\begin{pmatrix}
e^{i\phi}sin(\omega) & e^{i\phi}cos(\omega) \\
cos(\omega) & -sin(\omega)
\end{pmatrix}
\begin{pmatrix}
\hat{c}^\dagger_{out} \\
\hat{d}^\dagger_{out}
\end{pmatrix}
\label{eq:one}
\end{equation}

We first express the input state $\ket{11}$ as $\hat{a}^\dagger_{out}\hat{b}^\dagger_{out}\ket{vac}$ where $\ket{vac}$ is the vacuum state of the electromagnetic field. Transforming this state as per eq. B1, we get the output state 
\begin{equation}
\begin{aligned}
(e^{i\phi}sin(\omega)\hat{c}^\dagger_{out}+e^{i\phi}cos(\omega)\hat{d}^\dagger_{out})(cos(\omega)\hat{c}^\dagger_{out}-sin(\omega) \\\hat{d}^\dagger_{out})\ket{vac} = e^{i\phi}(\frac{sin(2\omega}{\sqrt{2}}\ket{20}- \frac{sin(2\omega}{\sqrt{2}}\ket{02} + \\ \frac{cos(2\omega}{\sqrt{2}}\ket{11})
\end{aligned}
\end{equation}

Following the same steps, it can be shown that the basis states $\ket{20}_{ab}$ and $\ket{02}_{ab}$ transform as follows.

\begin{equation}
\begin{aligned}
\ket{20}_{ab} = e^{2i\phi}(sin^2(\omega)\ket{20}_{cd} + cos^2(\omega)\ket{02}_{cd} + \\ \frac{sin(2\omega)}{\sqrt{2}}\ket{11}_{cd})
\end{aligned}
\end{equation}
\begin{equation}
\begin{aligned}
\ket{02}_{ab} = e^{2i\phi}(cos^2(\omega)\ket{20}_{cd} + sin^2(\omega)\ket{02}_{cd} - \\ \frac{sin(2\omega)}{\sqrt{2}}\ket{11}_{cd})
\end{aligned}
\end{equation}

If in eq. B1, we have $e + f = 0$ or in other words, $e=-f=g/\sqrt{2}$, and the input to the interferometer takes the form $\ket{\phi}_{ab} = d\ket{11}_{ab} + g\frac{\ket{20}_{ab}-\ket{02}_{ab}}{\sqrt{2}}$, the probability amplitude of $\ket{11}_{cd}$ at the output of the interferometer is

\begin{equation}
\begin{aligned}
s = e^{i\phi}(d \times cos(2\omega) + g \times sin(2\omega) cos(\phi))
\end{aligned}
\end{equation}

If $\phi=0$ and the phase difference between the complex probability amplitudes $d$ and $g$ is $\Delta$ the probability of finding the two input photons split into different output modes of the interferometer, i.e. the splitting efficiency S is
\begin{equation}
S = \frac{1}{2} + \frac{|d|^2 - |g|^2}{2} cos(4\omega) + \frac{2|d| |h| cos(\Delta)}{2} sin(4\omega)     
\end{equation}
Let $|d|^2-|g|^2=x$ and $2|d||h|cos(\Delta)=y$. Then the maximum value of the above expression for S is 
\begin{equation}
S_{max} = \frac{1}{2} + \frac{\sqrt{x^2+y^2}}{2}
\end{equation}
$S_{max}$ is 1, or in other words the splitting efficiency is a 100\%, when $x^2+y^2=1$, which occurs when $cos(\Delta)=1$. With this constraint, we get $x^2+y^2 = (|d|^2+|g|^2)^2$, which evaluates to 1 because of the normalization condition for the wavefunction (eq. B6). $cos(\Delta)=1$ is equivalent to saying that $d$ and $g$ are real. \\ 

\section{Derivation of splitting efficiency for entangled photons}

To generate a pair of entangled photons, we first initialize the system in figure 4 to state $\ket{e0g}$, with the three level atom in the excited state, feeder cavity containing zero photons and the two level atom in its ground state. We follow the same steps as in Appendix A1 to calculate the splitting efficiency with one difference. In this case, we need to calculate the probability of detecting photons conditioned on the event that the two photon pulse emitted by the three level atom enters the feeder cavity\cite{rosenblum2011photon}. Following these steps, we obtain the following expression for splitting efficiency
\begin{widetext}
\begin{equation}
S  = \frac{\splitfrac{32 \gamma ^2 (2 \gamma +2 \delta +3) \sin (4 \omega )+\left(-4 \gamma  \left(4 \gamma  \left(\gamma ^2+\gamma -2\right)+2 \gamma  (2 \gamma -3) \delta -5 \delta -7\right)-6 \delta -3\right) \cos (4 \omega )+}{4 \gamma  (2 \gamma  (2 \gamma +13) \delta +4 \gamma  (\gamma  (\gamma +5)+8)+19 \delta +17)+6 \delta +3}}{4 (2 \gamma +1)^2 (2 \gamma +3) (2 \gamma +2 \delta +1)}
\end{equation}

\end{widetext}

% whether or not they are actually referencetd in the text. This is appropriate
\nocite{*}

\bibliography{apssamp}% Produces the bibliography via BibTeX.

\end{document}